# Analyzing data citation practices using the *Data Citation Index*

**Nicolas Robinson-García, Evaristo Jiménez-Contreras**

EC3 Evaluación de la Ciencia y de la Documentación Científica, Universidad de Granada,

Colegio Máximo de Cartuja, s/n, Granada, 18071 (Spain)

EC3Metrics, Gran Vía de Colón, 48, 5º 10, Granada, 18071 (Spain)

Email addresses: elrobin@ugr.es; evaristo@ugr.es

**Daniel Torres-Salinas***

EC3Metrics, EC3 Evaluación de la Ciencia y de la Documentación Científica, Universidad de Navarra,

Gran Vía de Colón, 48, 5º 10, Granada, 18071 (Spain)

Email address: torressalinas@gmail.com

**Abstract**

We present an analysis of data citation practices based on the Data Citation Index from Thomson Reuters. This database launched in 2012 aims to link data sets and data studies with citations received from the other citation indexes. The DCI harvests citations to research data from papers indexed in the Web of Science. It relies on the information provided by the data repository as data citation practices are inconsistent or inexistent in many cases. The findings of this study show that data citation practices are far from common in most research fields. Some differences have been reported on the way researchers cite data: while in the areas of Science and Engineering & Technology data sets were the most cited, in Social Sciences and Arts & Humanities data studies play a greater role. A total of 88.1% of the records have received no citation, but some repositories show very low uncitedness rates. Although data citation practices are rare in most fields, they have expanded in disciplines such as crystallography and genomics. We conclude by emphasizing the role that the DCI could play in encouraging the consistent, standardized citation of research data – a role that would enhance their value as a means of following the research process from data collection to publication.

**Keywords:** Data Citation Index, data sharing, citation practices, scholarly communication, repositories

* To whom all correspondence should be addressed

**Introduction**

Lately we have witnessed a renewed interest in data sharing and the development of reproducible research (Anon, 2008; Peng, 2011). In the last few years researchers have been challenged with the management and processing of huge amounts of datasets for conducting large-scale studies in what is known as the 'Big Data' phenomenon (Lynch, 2008). These changes open new possibilities in all fields of scientific research, enriching the findings provided and broadening the scale of research studies (i.e., Spinney, 2012). Sharing research data at a large scale benefits funding bodies, as they see how their investment pays back through increased use and re-use of data (Wood et al., 2010). The research community also benefits as it facilitates meta-analyses based on previous research (Ramasamy et al., 2008) and improves the current peer review process, verifying and replicating results (Peng, 2011).

However, working with large amounts of data involves many changes in the way research is conducted, as well as on the infrastructure needed (Anon, 2008). On the one hand, data have to be made available for other researchers in a format and through open channels that allow reuse and reproducibility (Vision, 2010). On the other hand, this means that researchers should be willing to take the time and make the effort to share the data they produce, changing their habits and conduct, something which still is far from reality (Wallis, Rolando & Borgman, 2014).

Sharing data is costly in terms of time, infrastructure and funds (Tenopir et al., 2011). Although these practices are relatively common in some fields such as Genomics or Astronomy (Borgman, 2012), they are still rare in the scientific enterprise. In fact, in many cases researchers are not willing or not capable of facilitating access to their datasets after publishing a paper, although many journals require them to do so, if asked for (Savage & Vickers, 2009). Among other reasons, researchers refer to the time required, copyright restrictions, embargoes or lack of funding or recognition (Costas, Meijer, Zahedi & Wouters, 2012; Tenopir et al., 2011). Also, many researchers are unaware of much of the infrastructure or standards available for them to share data in a reusable manner (Arzberger et al., 2004).

If data sharing is to become a common practice, a change in the culture and research process will have to take place. It may protect against scientific fraud and improve the scientific method, but only if such data are managed and shared correctly (Doorn, Dillo & van Horik, 2013). One way of encouraging data sharing is by establishing a reward system by which researchers see a benefit to their efforts and the time invested. While data peer review may serve to validate the research data made available (Grootveld & van Egmond, 2012), citations would encourage data sharing, as they currently are the main yardstick of recognition used by researchers, funding bodies and journals to measure the performance of a paper, a research career or a journal. If one is to demonstrate the benefits of sharing data in terms of a positive citation effect, researchers may well consider adopting such practices. This is the line of argument used by Piwowar and colleagues (Piwowar, Day & Fridsma, 2007; Piwowar & Chapman, 2010; Piwowar & Vision, 2013). In their studies they have analyzed the citation effects of publications which share data, concluding that there is a positive relation between data sharing and citations.

A different approach to track data-related citations would be to monitor 'data citations' that is those directed not to publications which share data, but to the data sets themselves. In order to reference a given data set, researchers may adopt different approaches, citing either the original paper in which the data set was described, a data paper published in a journal, a data study or a data set. In this context, many tools are being developed in order to track the 'impact' of data such as Thomson Reuters' Data Citation Index (hereafter DCI). Also some data banks, such as Figshare, now include metrics such as views and are announcing the future inclusion of citations. Others, such as DataCite, are working on the standardization of data citation practices and providing DOIs to data sets.

All these tools consider research data as another 'published output'. This analogy allows us to presume that the same recognition system (citations) applies also to research data. However, such a 'metaphor' can be misleading (Parsons & Fox, 2013). Indeed, Mayernik (2012) points out that if we are to acknowledge the role of citations in regard to research data, we should review their meaning in this new context, as it may differ from 'ordinary' citations. Hence we should ask whether the motivations for citing data are the same as for citing research papers. Mooney and Newton (2012) report a lack of consistency in data references: omitting the source from which data was retrieved, authors' acknowledgment and lack of standards.

The DCI may well be a useful tool for the expansion of data citation practices and their standardization. This study focuses on the information provided by this database. Launched in 2012, it aims to solve four specific issues (Force & Robinson, 2014): 1) data access and discovery, by including in a single database references to research data spread through various institutional and disciplinary (data) repositories; 2) data citation, by adopting the DataCite standard and linking papers with data; 3) lack of willingness to deposit and cite data; and 4) lack of recognition and credit. So far, two studies have been reported analyzing or describing the DCI. Torres-Salinas, Martín-Martín and Fuente-Gutiérrez (2014) studied the coverage of the Data Citation Index (DCI) by fields, analyzing the number of repositories and the distribution of data sets and data studies by field. From their analysis they concluded that the DCI is heavily biased towards the Hard Sciences, the most common document type is data sets (94% of the total share) and four repositories represent around 75% of the database (Gene Expression Omnibus, UniProt Knowledgebase, PANGAEA and U.S. Census Bureau TIGER/Line Shapefiles). More recently, Force and Robinson (2014) described the selection process followed by Thomson Reuters for indexing repositories, the creation process of the structure of the database and records and data citation retrieval and linking with other citation indexes.

**Objectives of the study**

This paper presents a cross-disciplinary study of data citation practices based on the Data Citation Index. This new database represents a milestone in scholarly communication as it allows for first-hand observation of the development of citation practices related to research data in a similar vein to that presented by Garfield when he developed the Science Citation Index (Garfield, 1964). We focus on the DCI due to its uniqueness, a multidisciplinary database launched in October 2012 which indexes major data repositories (Thomson Reuters, 2012) and associates citation data to each record, providing the basis to develop data citation metrics (along with the rest of the Thomson Reuters' citation indexes). This study attempts to better understand and explain how common data citation practices are among fields, the forms of data which are more commonly cited, and the role of repositories as 'containers' of data sets and data studies.

This study builds upon preliminary results presented at the STI Conference 2014 held in Leiden, The Netherlands (Torres-Salinas, Jiménez-Contreras & Robinson-Garcia, 2014), deepening on data citation differences among fields and thoroughly discussing the findings in order to better understand the potential role data citation may have in order to foster data sharing practices. We present an analysis of the citation distribution of the Data Citation Index in order to assess on the relevance of the citation data contained in it.

Specifically, we aim at identifying different citation practices by broad areas, subject categories and repositories. The paper is structured as follows. In the section Material and methods we report on the data retrieval and processing and the construction of the broad areas analyzed. We also offer a brief description of the fields available in each record in the database and the document types which will serve to discuss our findings. These are presented in the section Results. In particular, we analyze the distribution of records and citations by document type and area. We present a longitudinal analysis of the citation distribution. We analyze the number of citations by subject categories and repositories. Finally we reflect on the potential of this type of analysis as well as of the DCI as a data source for conducting them.

**Material and methods**

In this paper we conduct an analysis of the citation distribution of the Data Citation Index by areas and repositories. Between May and June, 2013, we retrieved all records indexed in the DCI and created a relational database for data processing. Subject categories to which repositories were assigned were aggregated into four broad areas (Science, Engineering & Technology, Social Sciences and Arts & Humanities). We applied full counting – records assigned to more than one area were included in both. For more information related to the construction of these areas, the reader is referred to Robinson-Garcia, Jiménez-Contreras and Torres-Salinas (2014).

The DCI is included within the Web of Science platform, with the Web of Science Core Collection (which includes the SCI, SSCI and A&HCI) and other of the databases offered by Thomson Reuters. The DCI follows a selection process in order to maintain certain standards of quality. The criteria followed for indexing repositories include factors such as publication standards, editorial content, international diversity of authorship, geographic origin and scope or citation data associated with it (Thomson Reuters, 2012).

Regarding how data are cited and linked in the DCI, Thomson Reuters states that it has adopted and encourages the DataCite citation standard (Swoger, 2012), by which citations should include at least the following elements (Starr & Gastl, 2011): 1) an identifier (currently it employs DOIs), 2) a creator/s (researchers responsible for producing the data or the publication), 3) title of the data set or data study, 4) publisher (defined as the place where the data is deposited), and 5) publication year (indicating when was the data made publicly available). But not all data sets and studies include a DOI (depending on the originating repository), hindering the ability to establish links.

Currently, the DCI relies on the information provided by the data repository regarding publications in which the data set or data study was cited. In figure 1 we show an example of how this link is performed. As observed, the citing paper does not 'cite' the data study but mentions it. The citation is included in the record extracted from the repository where the data study is deposited. Hence, the citations are provided by the repositories themselves.

Figure 1. An example of how the Data Citation Index links data sets and studies with publications.

**1. Cited data study**

**2. Citing paper**

**3. Link retrieved from the repository**

The DCI includes three different document types: data sets, data studies and repositories (Figure 2). According to Thomson Reuters (2012), data repositories are defined as databases which store and provide access to the raw data contained in data sets and data studies. Data sets are single and coherent sets of data provided as part of a collection, data study or experiment in one or a more files (Thomson Reuters, 2012). Finally, data studies are defined as a description of experiments with associated data which have been used in these experiments (Thomson Reuters, 2012). All data sets and data studies are assigned to a repository, serving the latter as a container of research data in the same vein as journals contain articles. However, repositories also receive citations and are therefore included as document type. Data sets are single files of data lacking any description of the data set other than the abstract. In many cases, data sets may be linked to data studies; hence these may contain several data sets as well as the description of the data collection and processing. The distribution of each document type varies by repository. While some repositories include both datasets and data studies (i.e.,

PANGAEA), others only include one of them (i.e., Animal QTL Database for data sets only and UK Data Archive for data studies only). Also, not all fields in records seem to be common to all repositories, following instead different structures depending on the repository from which data was retrieved.

Figure 2. Screenshot of the results page of the Data Citation Index. Filter for document types is highlighted

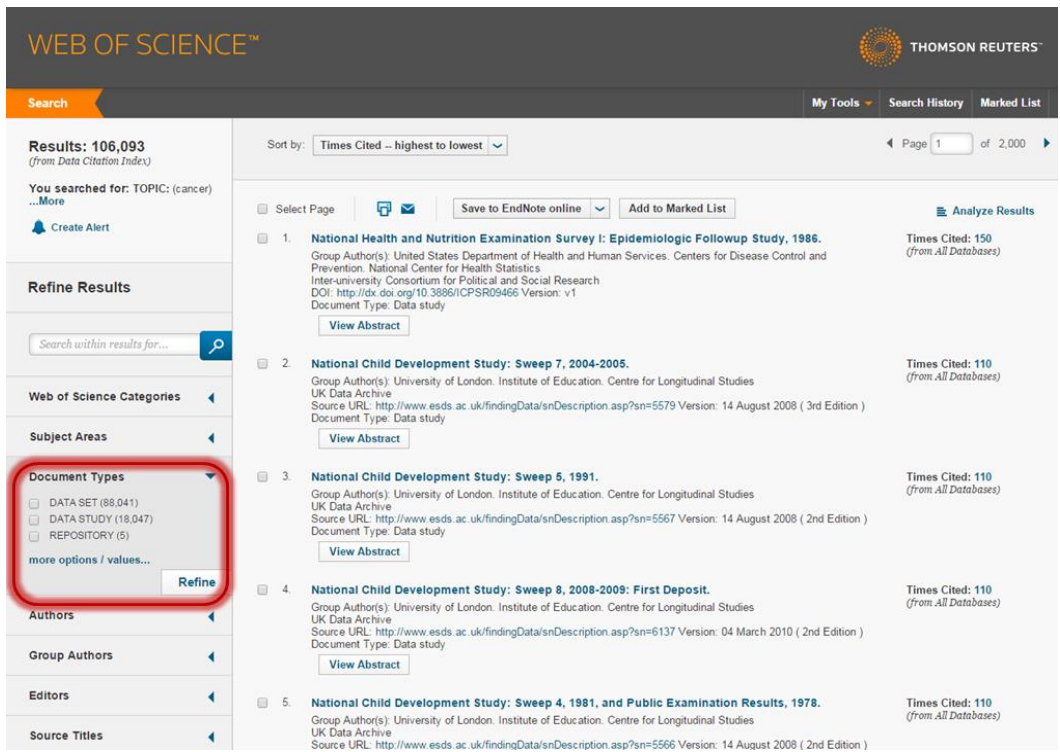

**Results**

In order to provide a comprehensive description of the database as well as to explore data citation practices among fields, this section has been structured as follows. First we show an overall view of the database and a temporal evolution of the citation and records distribution. Then, we show data citation differences as well as coverage limitations between four broad fields: Science, Engineering & Technology, Social Sciences and Arts & Humanities. We focus on subject categories, as a means to deepen into such differences. Finally we descend to the repository level, analyzing the role played by the repositories which include the highest number of citation in the DCI.

*General overview*

There were a total of 2,626,528 records in the DCI by retrieval date of May, 2013 (table 1). Most of these are data sets, representing 94% of the database. Regarding the citation distribution, 88.1% of all records remain uncited. Data studies receive more citations on average (0.69) than data sets (0.12), but again, data sets accumulate most of the citations included in the DCI (73%).

Table 1. Indicators for all records, repositories, data sets and data studies

|  | All Document Types | Data repositories | Data studies | Data sets |
|---|---|---|---|---|
| Total Citations | 404,211 | 3,265 | 106,895 | 294,051 |
| Total Records | 2,623,528 | 90 | 154,674 | 2,468,736 |
| Uncited Records | 2,311,553 | 63 | 126,428 | 2,185,062 |
| % Uncited | 88.11 | 40.0 | 81.74 | 88.51 |
| Citation Average | 0.15 | 36.28 | 0.69 | 0.12 |
| Standard Deviation | 3.06 | 336.07 | 9.56 | 0.36 |

The DCI tracks data sets back to 1800 and data studies back to 1865. The earliest data sets found belong to the UK Data Archive and are derived from the British Geological Survey, containing miscellaneous geological information from various areas of Great Britain (i.e., http://data.gov.uk/dataset/50k-sheet-data-files). The earliest data study was found in The Association of Religion Data Archives and contains New York censuses from 1855 to 1865 with social, political and economic indicators from every town and city of the state of New York (http://www.thearda.com/Archive/Files/Descriptions/NY185565.asp).

99.7% of the records were published in the period 1951-2013. In figure 3 we show the evolution of records and citation for data sets and data studies according to the DCI in this period. We excluded 2013 from the figure as this year was incomplete at the time of the data retrieval. As observed, the increase within the 1951-2012 time period has been exponential, with an average annual relative growth rate (ARG) of 8.7% for data sets and 10.1% for data studies in this time period. The ARG for the last decade is of 24% for data sets and 12.1% for data studies.

Also an exponential growth of citations to data sets can be observed since 1951, but not to data studies, which show a more irregular pattern. Despite this, citations to both document types reach similar figures at the end of the analyzed period.

Figure 3. Longitudinal evolution of citations and records for data sets and data studies according to the Data Citation Index. Time period 1950-2012

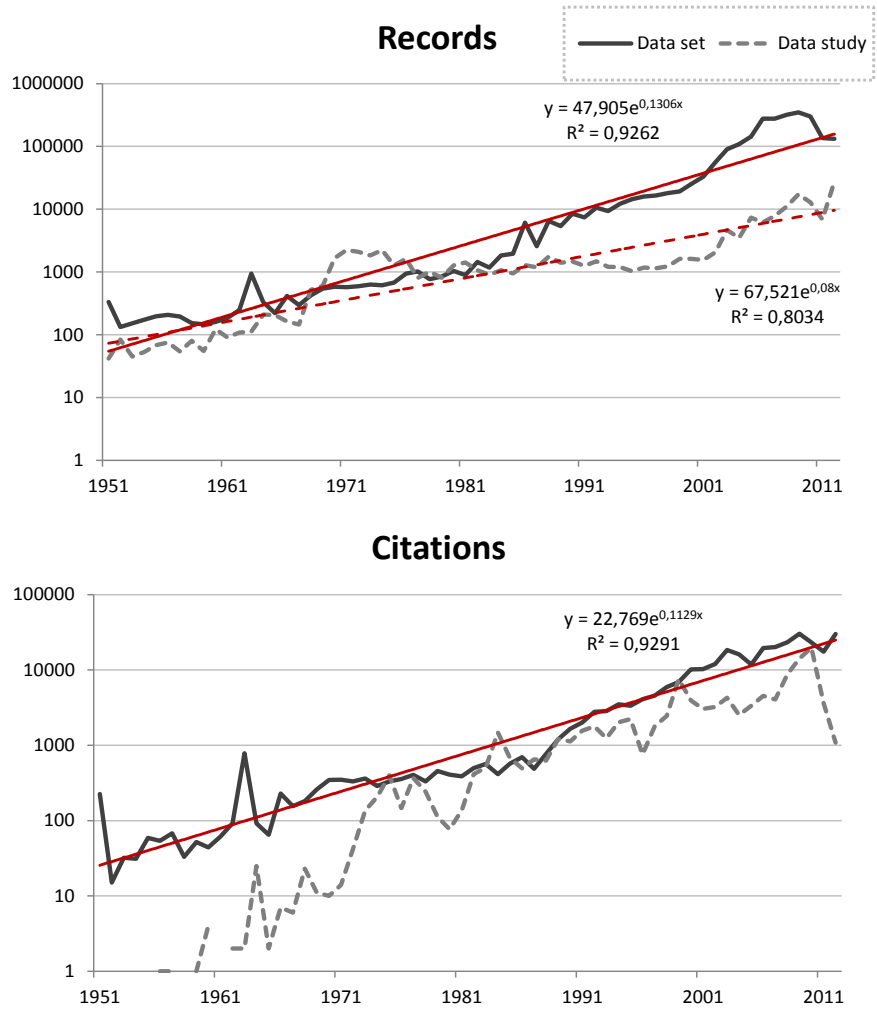

*Analysis by areas*

Table 2 analyzes the number of records and citations by areas. A total of 81% of the records belong to the area of Science, followed far behind by Social Sciences (18%). On the other hand, Engineering & Technology is the most underrepresented area with 0.1% of the whole share. This pattern is also seen when focusing on data sets only. Science represents 81% of the database followed by Social Sciences with a share of 17%. The picture changes slightly when focusing on data studies. Although the distribution is still severely biased towards Science (74%), Social Sciences have a higher presence (24%).

Regarding the citation distribution, only in the area of Engineering & Technology do we see a citation average above 0.5, highlighting the high degree of uncitedness (share of records receiving no citations). Indeed, the high standard deviation values mean that the average number needs to be interpreted with great care. Science accumulates most citations (79%) followed by the Social Sciences (18%), Arts & Humanities (5%) and finally, Engineering & Technology (0.2%)., but there are significant differences by document types. Although in the fields of Engineering & Technology and Science, researchers tend to cite data sets (97% of all citations received in Engineering & Technology and 92% in Science are directed to data sets), the opposite occurs in Social Sciences and Arts & Humanities, where most of the citations were directed to data studies (96% in the case of the former and all except one citation in the case of the latter).

Table 2. Indicators for all records, datasets and data studies by area

A. All document types

|  | Total Records | % Records | Total Citations | % Citations | Citation Average | Standard Deviation |
|---|---|---|---|---|---|---|
| Engineering & Technology | 1,786 | 0.07 | 916 | 0.23 | 0.51 | 0.90 |
| Humanities & Arts | 51,444 | 1.96 | 20,460 | 5.06 | 0.40 | 7.99 |
| Science | 2,118,855 | 80.76 | 319,458 | 79.03 | 0.15 | 0.59 |
| Social Sciences | 462,826 | 17.64 | 72,855 | 18.02 | 0.16 | 6.84 |

B. Datasets

|  | Total Records | % Records | Total Citations | % Citations | Citation Average | Standard Deviation |
|---|---|---|---|---|---|---|
| Engineering & Technology | 1,545 | 0.06 | 890 | 0.30 | 0.58 | 0.94 |
| Humanities & Arts | 44,588 | 1.81 | 1 | 0.00 | 0.00 | 0.00 |
| Science | 2,004,449 | 81.19 | 293,193 | 99.71 | 0.15 | 0.40 |
| Social Sciences | 424,952 | 17.21 | 7 | 0.00 | 0.00 | 0.01 |

C. Data studies

|  | Total Records | % Records | Total Citations | % Citations | Citation Average | Standard Deviation |
|---|---|---|---|---|---|---|
| Engineering & Technology | 240 | 0.16 | 26 | 0.02 | 0.11 | 0.50 |
| Humanities & Arts | 6,847 | 4.43 | 20,459 | 19.14 | 2.99 | 21.72 |
| Science | 114,338 | 73.92 | 26,189 | 24.50 | 0.23 | 1.91 |
| Social Sciences | 37,855 | 24.47 | 69,659 | 65.17 | 1.84 | 17.34 |

*Citation analysis by subject categories*

The bias towards the area of science is later confirmed when analyzing the citation distribution by subject categories. Figure 4 shows the top 10 subject categories according to the DCI with a higher number of citations received by document type. All top ten subject categories for data sets receiving citations belong to the area of Science. Also, we observe that a single subject category, Crystallography, accumulates almost half of all citations to data sets. This category along with Biochemistry & Molecular Biology and Genetics & Heredity represent 86% of all citations. However, while there are 1,224,247 data sets in Biochemistry & Molecular Biology, only 152,235 data sets are assigned to the subject category of Crystallography. The field of Physics, Atomic, Molecular & Chemical has the highest citation average (0.98, std. dev. 0.15), followed by Medical Laboratory Technology (0.96, std. dev. 0.32) and Nanoscience & Nanotechnology (0.96, std. dev. 0.32).

Figure 4. Distribution of citations received by document type for the top 10 most highly cited subject categories according to the Data Citation Index

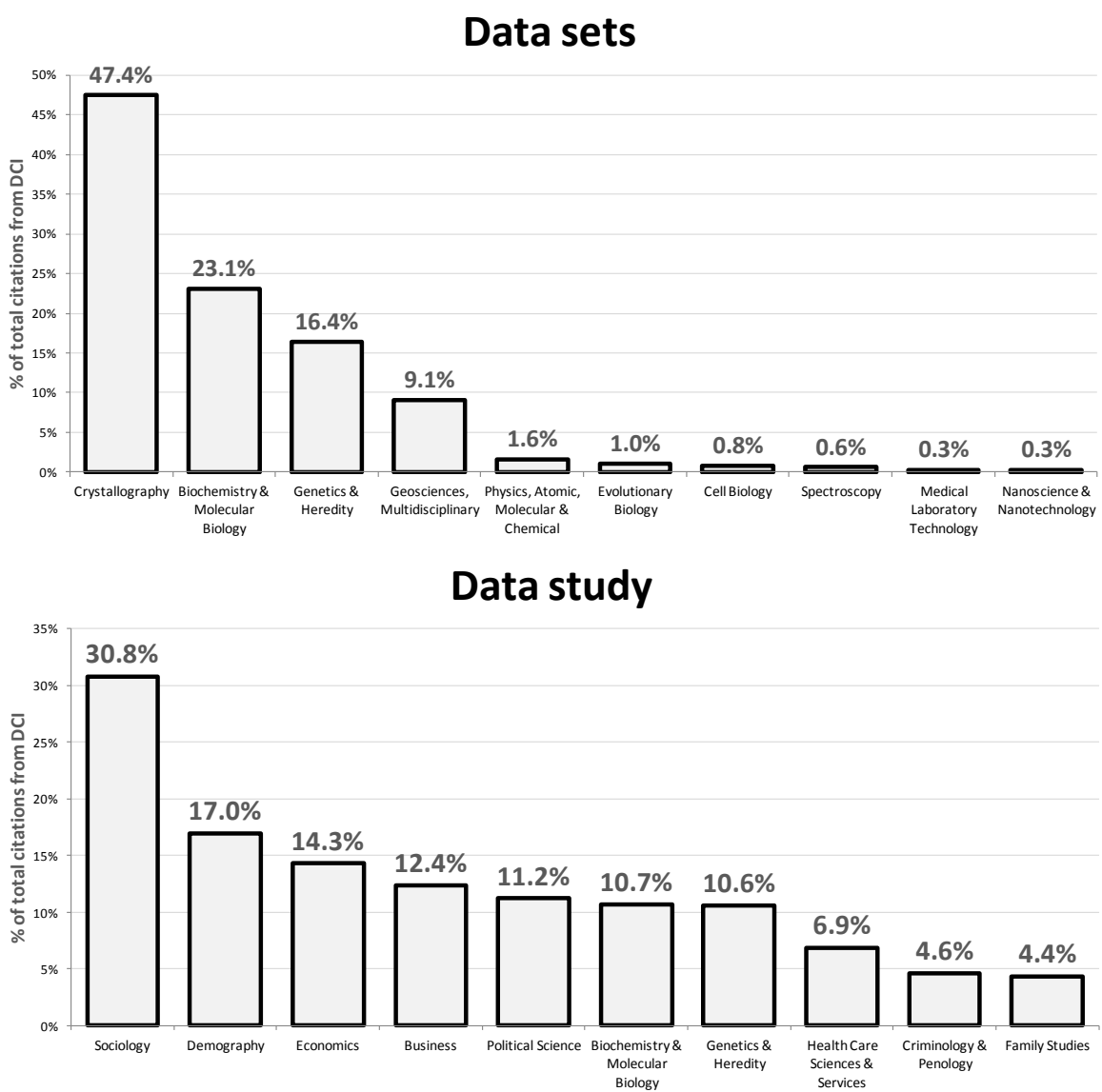

The pictures changes radically in the case of data studies. Here, seven of the top ten categories belong to the area of Social Sciences. Sociology accumulates 30.8% of all citations directed at data studies in the DCI, followed by far by Demography (17.0%) and Economics (14.3%). In this case, Sociology is the third largest subject category of the ten (20,438 records), only behind of Genetics &

Heredity (61,023) and Biochemistry & Molecular Biology (23,425). Health Care Sciences & Services has the highest citation average (9.05, std. dev. 56.9), followed by Business (5.63, std. dev. 24.79) and Demograpy (4.74, std. dev. 28.35).

Figure 5. Top 5 subject categories with the highest number citations received by data sets in the Data Citation Index.

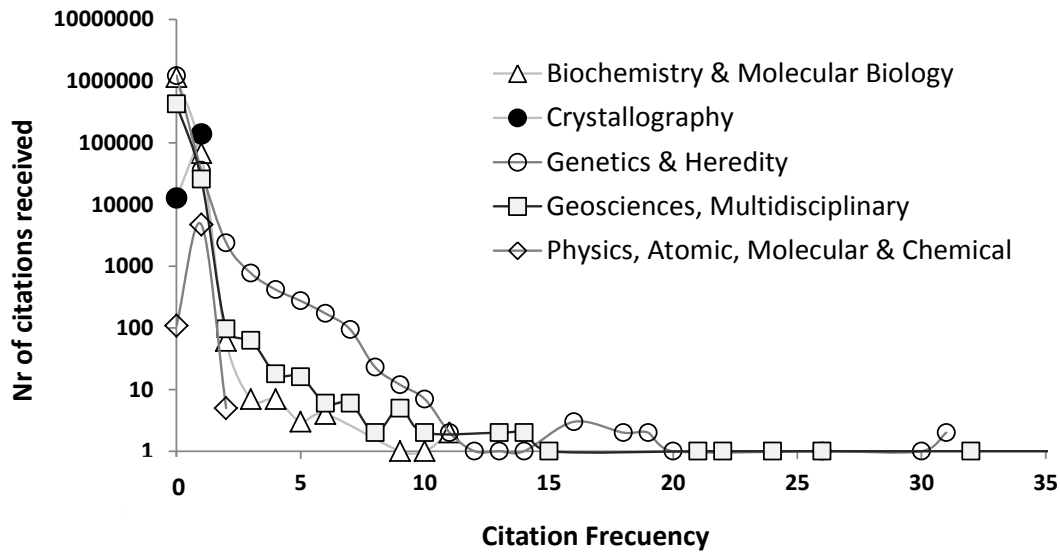

Figures 5 and 6 show the distribution of the top 5 subject categories displayed in figure 2 by document type. The x-axis shows the number of records, while the y-axis shows the number of citations received. The y-axis is in logarithmical scale. For the sake of clarity we omit data regarding the rest of the subject categories; this information is available at Robinson-Garcia, Jiménez-Contreras and Torres-Salinas (2014). In figure 5 we observe that the citation distribution for datasets in Biochemistry & Molecular Biology, Genetics & Heredity, Geosciences, Multidisciplinary, and Physics, Atomic, Molecular & Chemical is highly skewed, with most of the records having received zero or one citation. The only exception can be found in the subject of Crystallography, which shows a rare distribution, as all cited records have received only one citation.

Figure 6 shows a completely different pattern for data studies. Citation distributions are not as skewed and the pattern is more irregular than in the case of data sets. Also here, differences between subject categories are not as significant as in the case of data sets.

Figure 6. Top 5 subject categories with the highest number citations received by data studies in the Data Citation Index.

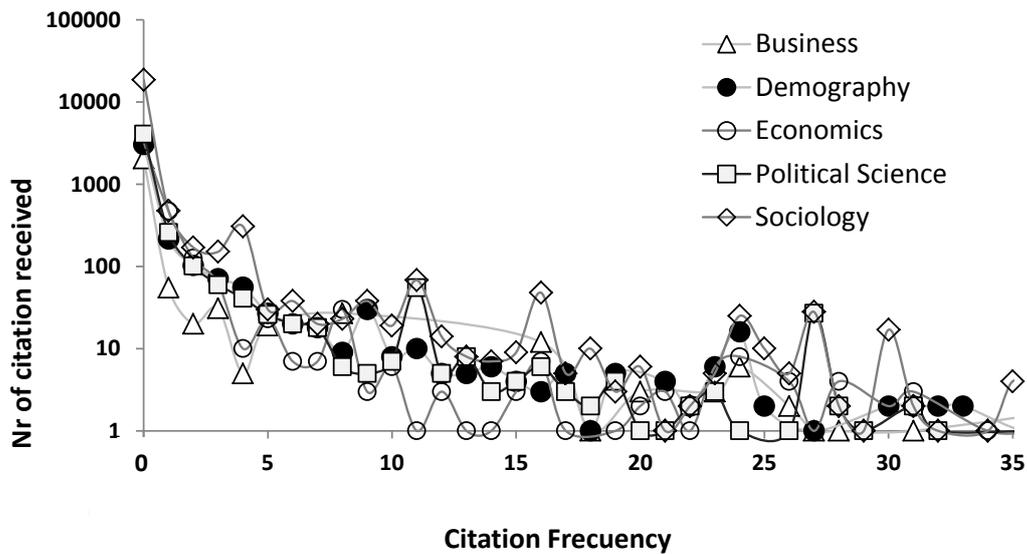

*Relation between citations, number of records and repositories*

Next we focus on repositories and their relation with records and citations. This will allow us to see whether data sharing and citing practices are more common in certain fields than others. Also, we can determine whether some repositories are more used for depositing and citing data, playing a greater role within a given field. In order to explore this, in figure 7 we relate the number of records with the number of citations received for the largest repositories indexed in the DCI.

Here we see that the repository with more citations is specialized in Crystallography (Crystallography Open Database), followed by the Protein Data Bank (Biochemistry & Molecular Biology) and the Inter-university Consortium for Political and Social Research (Social Sciences, Interdisciplinary. Also, these three repositories are the ones containing a higher number of citations. The share of cited records in each repository varies greatly while most records are cited in the Crystallography Open Database, the opposite occurs in PANGAEA and the Gene Expression Omnibus. Most repositories have a low number of records and citations, though in some cases the share of cited records is very high (see e.g., MiRBase or Animal QTL Database in Figure 7). Indeed, we observe that 43 repositories (Robinson-Garcia, Jiménez-Contreras & Torres-Salinas, 2014) have no citations at all, meaning that either these are not indexed by the repository and hence, not included in the DCI, or that they belong to fields where data citation does not take place.

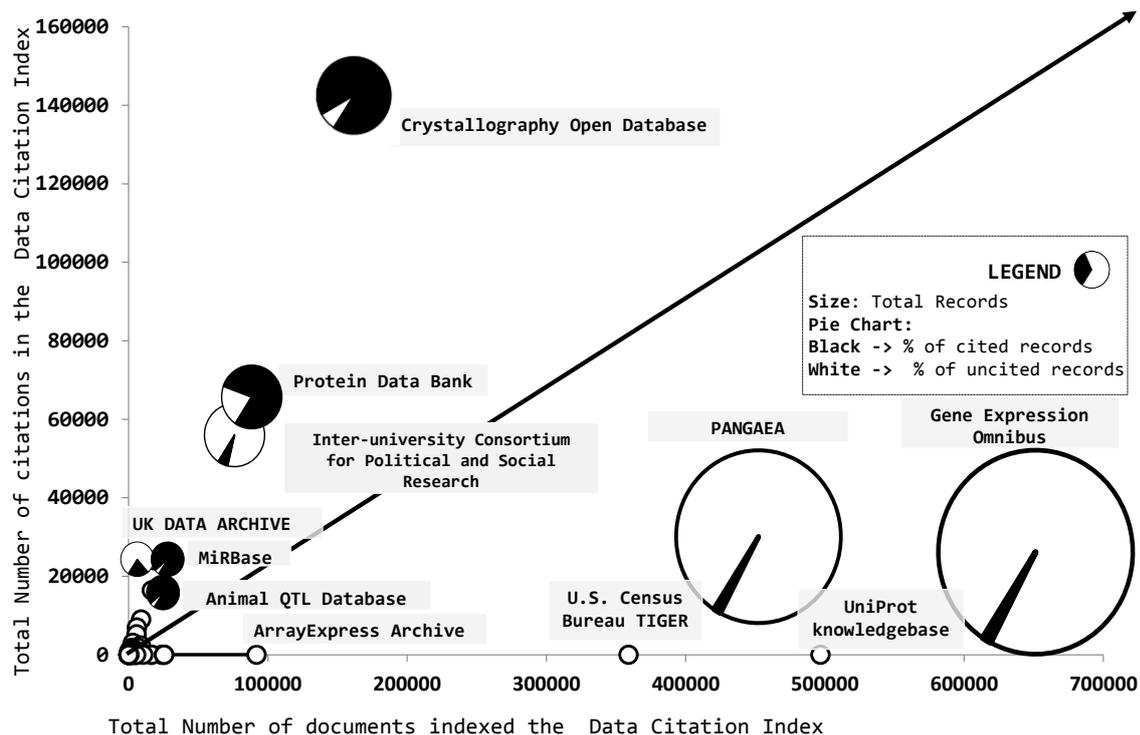

Figure 7. Main repositories in the DCI, citations received and total number of records

Table 3 shows the ten largest repositories by document type. As observed, citations are quite spread across some repositories. Here we observe how two repositories (Crystallography Open Database and Protein Data Bank) which barely represent 10% of the data sets included in the DCI account for 67.9% of the total number of citations received. In the case of data studies, the same happens with the Inter-university Consortium for Political and Social Research and the UK Data Archive, which represent 8.3% of the total data studies included in the database but concentrate 75.6% of the total citations directed at this document type.

We also observe that the standard deviation of the citation distribution in top cited repositories is not as high. This is reinforced by the low uncitedness rate in some repositories, which is far from the figures presented for the overall database (see table 1). In fact, in the case of data sets, only two of the ten top repositories have an uncitedness rate higher than 70% while the rest are around or under 10%. In the case of data studies, these rates are higher. Here we observe that the highest uncitedness rate goes to the Australian Data Archive (90.8%) followed by the UK Data Archive (79.4%) and GWES Central (76.6%).

Table 3. Output and citation indicators for top ten repositories with a higher number of citations in the Data Citation Index according to their document type

A. Data sets

| Repository | Citations | | | | Records | | |
|---|---|---|---|---|---|---|---|
| | Total | Avg* | Std dev* | % DCI* | Total | %Uncited* | % DCI |
| Crystallography Open Database | 139434 | 0.92 | 0.27 | 47.42 | 150916 | 7.61 | 6.11 |
| Protein Data Bank | 60197 | 0.79 | 0.41 | 20.47 | 76562 | 21.37 | 3.10 |
| PANGAEA | 25468 | 0.06 | 0.23 | 8.66 | 443088 | 94.25 | 17.95 |
| miRBase | 24092 | 1.32 | 1.02 | 8.19 | 18221 | 4.48 | 0.74 |
| Animal QTL Database | 16456 | 0.99 | 0.10 | 5.60 | 16635 | 1.08 | 0.67 |
| Cancer Models Database | 6972 | 1.17 | 1.31 | 2.37 | 5934 | 10.13 | 0.24 |
| Nucleic Acid Database | 5232 | 0.94 | 0.25 | 1.78 | 5595 | 6.49 | 0.23 |
| nmrshiftdb2 | 4736 | 0.98 | 0.15 | 1.61 | 4839 | 2.23 | 0.20 |
| TreeBASE | 3056 | 1.00 | 0.00 | 1.04 | 3056 | 0.00 | 0.12 |
| The Cell: An Image Library | 2365 | 0.27 | 0.47 | 0.80 | 8788 | 74.00 | 0.36 |

B. Data studies

| Repository | Citations | | | | Records | | |
|---|---|---|---|---|---|---|---|
| | Total | Avg | Std dev | % DCI | Total | %Uncited | % DCI |
| Inter-university Consortium for Political and Social Research | 55271 | 6.98 | 37.56 | 51.71 | 7919 | 49.68 | 5.12 |
| UK Data Archive | 25588 | 5.15 | 22.19 | 23.94 | 4964 | 79.43 | 3.21 |
| Gene Expression Omnibus | 10264 | 0.46 | 0.54 | 9.60 | 22138 | 55.40 | 14.31 |
| nmrshiftdb2 | 4259 | 1.03 | 0.28 | 3.98 | 4122 | 2.09 | 2.66 |
| PANGAEA | 3728 | 0.92 | 0.27 | 3.49 | 4048 | 7.95 | 2.62 |
| Dryad | 1847 | 0.86 | 0.35 | 1.73 | 2152 | 14.27 | 1.39 |
| Finnish Social Science Data Archive | 1137 | 1.38 | 2.82 | 1.06 | 824 | 50.61 | 0.53 |
| Electron Microscopy Data Bank | 1109 | 0.88 | 0.37 | 1.04 | 1262 | 13.71 | 0.82 |
| Australian Data Archive | 857 | 0.41 | 2.96 | 0.80 | 2106 | 90.79 | 1.36 |
| GWAS Central | 646 | 0.23 | 0.42 | 0.60 | 2762 | 76.61 | 1.79 |

**Note:** %Uncited: Percentage of uncitedness %DCI: Percentage of records/citations from the total share in the database Avg: Average citations per record Std dev: standard deviation

**Discussion and concluding remarks**

The DCI represents a milestone in the development of an environment which facilitates searching, retrieving and following research from data to publication and the different products arising (data sets, studies, articles, etc.). The DCI is still in its infancy as the 'data citation culture' is not yet well established among researchers. However, the DCI may stimulate sharing data and standardization in citing it, in the same way that the Science Citation Index did with referencing research papers (Garfield, 1970). The findings of this study confirm that data citation practices are far from common within the scientific community, with a high rate of uncitedness (88%). There seem to be different citation practices: while researchers from the areas of Science and Engineering & Technology cite data sets, in Social Sciences and Arts & Humanities data studies are more cited (table 2). This fact is important, as it will determine how citation and publication analyses should be designed when analyzing data sharing practices; the chosen field will determine the suitability of one document type or the other.

When focusing on specific disciplines, we observe that a single repository, Crystallography Open Database, which represents 6.11% of all records included in the DCI, accounts for 47.42% of all citations included (table 3). While citation distributions for data sets and data studies are skewed (figures 5 and 6), in the case of Crystallography, records have either one citation or no citations (figure 5). This field has a long-standing tradition on data sharing since 1971 (Cech et al., 2003) as well as their own standard for information interchange, the Crystallography Information Framework (CIF). In this sense, the reason for such a strange distribution may be that most citations are self-citations. If this is the case, two interpretations are plausible. Either data sets are only cited by their producers or the repositories in this discipline have not captured all citations other than the one from the original paper. Further research is necessary to confirm this point.

This paper presents a cross-disciplinary analysis of data citation practices and differences among fields based on the Data Citation Index. Data sharing is starting to be seriously encouraged by many funding bodies and research organizations. Although these practices bring theoretically obvious benefits to the research community, lack of awareness, cost and the effort required to do so constitute serious drawbacks. If recognized by the community, researchers may feel encouraged and undertake the necessary efforts to share their scientific data. In this sense, citations may be a way of recognition (Costas et al., 2012). However, little is known about data citation practices (Costas, et al., 2012; Parsons et al., 2010; Tenopir, et al., 2011). Although some citation standards have been developed (Starr & Gastl, 2011), researchers are not consistent when referring to data sets, often simply mentioning them. There is also no common practice when publishing data (Costas et al., 2012), even if depositing data in a data bank could be considered equivalent to publishing (Parsons & Fox, 2013). In this early stage, data repositories play a crucial role linking data sets and data studies with publications, and here is where the DCI may encourage consistent citing of research data.

There are also important questions that need to be raised. Is data citing the same as data sharing? Citation practices are not common to all areas of scientific knowledge and only certain fields have developed an infrastructure that allows researchers to use and share data, but still the link between sharing and citing is missed. When focusing on the top repositories which included a higher number of citations we observed that the uncitedness rate varies greatly among repositories. This shows that data citation practices may be well developed in some fields. The concentration of citations in a small number of repositories also raises the question of the suitability of the repositories indexed in the DCI, as many of them have no citations at all. This could be due not to poor choice, but to the difficulty of linking references to data sets and data studies. If the citation links are provided by the repositories themselves (as shown in the method section) this may limit the use of the DCI. In this regard, we observe that the DCI is heavily biased towards certain fields from the Hard Sciences (Torres-Salinas, Martín-Martín & Fuente-Gutiérrez, 2014), with almost no representation from Engineering & Technology. The reasons for this may not only be attributed to the criteria adhered to by Thomson Reuters, but again to the expansion of data citation practices within the research community.

Data citation analysis may encourage researchers to make their data publicly available as they will be able to analyze the impact of their contribution and the use of their work as well as developing a more open and transparent research process. Other repositories of a multidisciplinary nature have been launched in the recent years such as Figshare (http://figshare.com) which also seek at including metrics that will indicate the use and discussion awakened by the data displayed. However, as data citation practices develop, more analyses will be necessary with regard to document type. In this article we focused on data sets and data studies present in the DCI, but other types should be considered, such as data papers for instance. As Garfield stated: '"Perfection" through citation indexing may not be practical for several years, but our present efforts appear quite satisfactory for the costs involved and the results achieved' (Garfield, 1983).

**Supplementary Material**

Detailed data on the construction of the four broad fields along with supplementary tables to figures 5, 6 and 7 are available at Robinson-Garcia, Jiménez-Contreras and Torres-Salinas (2014).

**Acknowledgments**

Thanks are due to the two anonymous referees for their helpful comments. The authors would also like to thank Ismael Rafols and Inma Aleixos for fruitful discussions regarding data sharing practices in the field of Crystallography.